\documentclass[preprint]{revtex4}

\usepackage{graphicx,psfrag}
\usepackage{amssymb,amsmath,amsthm}
\usepackage{verbatim}
\usepackage{color}

\definecolor{magenta}{rgb}{0.7,0,0.7}
\definecolor{orange}{rgb}{1,0.5,0}

\newcommand{\beq}{\begin{equation}}
\newcommand{\eeq}{\end{equation}}
\newcommand{\bea}{\begin{eqnarray}}
\newcommand{\eea}{\end{eqnarray}}

\begin{document}
\def\Fbox#1{\vskip1ex\hbox to 8.5cm{\hfil\fboxsep0.3cm\fbox{%
\parbox{8.0cm}{#1}}\hfil}\vskip1ex\noindent}  
\newcommand{\B}[1]{{\bm{#1}}}
\newcommand{\C}[1]{{\mathcal{#1}}}
\renewcommand{\it}[1]{\textit{#1}} \newcommand{\Onecol}
{\begin{widetext} \onecolumngrid} 
\newcommand{\Twocol} {\end{widetext} \twocolumngrid}   

\begin{abstract}
Dislocations are at the heart of the plastic
behavior of crystalline materials yet it is notoriously
difficult to perform quantitative, non-intrusive, measurements
of their single or collective properties. Dislocation density is a critical variable that determines dislocation mobility, strength and ductility. On the one hand, individual dislocations can be probed in detail with transmission electron microscopy. On the other hand, their collective properties must be simulated numerically. Here we show that ultrasound technology
can be used to measure dislocation density. This
development rests on theory---a generalization of the
Granato-L\"ucke theory for the interaction of elastic waves
with dislocations---and Resonant Ultrasound Spectroscopy (RUS)
measurements. The chosen material is aluminum, to which different dislocation contents were induced through annealing and cold rolling processes. The dislocation densities obtained with RUS compare favorably with those inferred from X-ray diffraction, using the modified Williamson-Hall method.

\end{abstract}
\title{Ultrasound as a probe of dislocation density in aluminum}
\author{Nicol\'as Mujica$^{1,2}$, Mar\'\i a Teresa Cerda$^1$, Rodrigo Espinoza$^3$, \\ Judit Lisoni$^{1}$ and Fernando Lund$^{1,2,*}$  }
\affiliation{$^1$Departamento de F\'isica and $^2$CIMAT, Facultad de Ciencias F\'\i sicas y Matem\'aticas, \\ Universidad de Chile,
Avenida Blanco Encalada 2008, Santiago, Chile.}
\affiliation{$^3$Departamento de Ciencia de los Materiales, Facultad de Ciencias F\'\i sicas y Matem\'aticas, \\ Universidad de Chile,
Avenida Tupper 2069, Santiago, Chile.}

%
\maketitle
\section{Introduction}
First introduced as a curiosity in the mathematical description of elastic continua, dislocations have become a fundamental building block  in the explanation of the plastic behavior of crystalline materials. The advent of transmission electron microscopy (TEM) enabled their direct visualization and provided a firm experimental foundation for their subsequent study. Nowadays, dislocations not only describe a physical reality, they also provide a conceptual framework to explain very many phenomena in crystal plasticity. However,  quantitative modeling with predictive power has remained elusive. 

At the atomistic level, there has been significant progress in the theoretical, direct observation, and numerical modeling of individual dislocations. A few examples are the work of Leyton et al. \cite{curtin} on the prediction of energy barriers to dislocation motion,  the direct observation  of dislocation mechanisms of Oh et al. \cite{shoh},  and the molecular dynamics simulations of Marian et al. \cite{marian}.  However, in spite of the considerable advances in the study of individual dislocations, the fact remains that the collective behavior of many dislocations, a critical ingredient in crystal plasticity, is not necessarily obtainable by simple aggregation of the individual behavior of many isolated dislocations. In fatigue, for example, a non-intrusive characterization of the material prior to fatigue failure would represent a significant step towards the control of fatigue damage. Progress in this respect has been achieved through the use of infrared thermography \cite{fat1,fat2,fat3}.  On the numerical modelling front, a direct approach to tackle the problem of the collective behavior of dislocations is the Dislocation Dynamics Method \cite{Kub97,Sch99,Gho99,Cai04}. In a related development, Warner et al. \cite{curtin2} have performed a multiscale simulation that sheds light on the influence of dislocation formation on the propagation of cracks in aluminum.  

Dislocation density is closely related to the ability of a given material to endure plastic deformation \cite{dd1,dd2,dd3}. But, can dislocation density be accurately measured? TEM allows for a local measurement in very small specimens that have to be especially prepared, including a non direct thickness measurement to estimate the local volume of the area under observation \cite{tem}. This technique is of limited use when individual dislocations cannot be resolved due to high densities \cite{balogh}. Another widely used possibility is to consider the broadening of X-ray diffraction (XRD) peaks by dislocations and other defects \cite{re9}, a formulation that has been applied to submicron grain size and heavily deformed copper \cite{re9, re10, re11, re12}, nanocrystalline iron powder and iron-based powders \cite{re13,re14,s2}, and aluminum alloys \cite{re15, re16, re17}. Recently, Balogh, Capolungo and Tom\'e \cite{balogh} have provided an assessment of this method, including its strengths and limitations. Clearly, an alternative, reliable, non-intrusive, way to measure dislocation density would appear to be a desirable development.  

Ultrasound (US) technology has been widely used as a non-destructive evaluation tool for several decades \cite{NDE}. This is based on the fact that the associated ultrasonic energy is very low compared to energies needed to deform, break, or significantly alter the tested material in any way and on the further fact that interfaces, as well as flaws, within a material, affect the propagation of acoustic waves.  So, monitoring ultrasound transmission or reflection provides information about the insides of a material or structure. 

Dislocations interact with elastic waves. This has been known since the early days of dislocation theory \cite{nabarro,eshelby}. Can ultrasound technology be used to learn something---quantitatively but non-intrusively---about dislocations in a crystalline solid? The effect that dislocations have on nonlinear wave propagation has been studied \cite{nl1,nl2}, as well as the possibility of using it to characterize fatigue microstructures \cite{nl3}. More specifically, in recent years Resonant Ultrasound Spectroscopy (RUS) has emerged as a very efficient tool for materials characterization \cite{migliori93,leisure97,schwarz2000,radovic2004,rusbook}.  A complete set of elastic constants can be measured precisely for any kind of symmetry. It has proved to be useful in different fields, as in condensed matter where its precision has been essential in order to demonstrate that a quasi-periodic lattice is indeed elastically isotropic \cite{spoor95}. Other examples are the observation of giant softening in high temperature superconductors \cite{migliori90}, the measurement of the mechanical properties of  biomaterials \cite{kinney2004}, the determination of elastic constants of thin films \cite{Ogi_thin}, the mechanical characterization of composite materials \cite{Ogi_composite}, the measurement of internal friction in polycrystalline copper \cite{ledbetter_friction}, the characterization of elastic anisotropy in a bulk metallic glass \cite{greer2011}, and its application in geophysics, in particular for the measurement of elastic constants at high temperature \cite{PhysTod}. A recent publication \cite{radovic2004} compares RUS to other more traditional techniques, including four-point bending, nanoindentation and impulse excitation, concluding that it is much more precise than most of them for the measurement of elastic constants. 

In this article we show that, other things being equal, RUS can distinguish between materials with different dislocation densities. This is demonstrated preparing samples with different dislocation content from the same as-received aluminum bar (Section \ref{sec2}), and measuring their elastic constants and elastic wave velocity with RUS (Section \ref{sec3}). This provides a measure of their dislocation density (Section \ref{sec4}). The results are further compared and corroborated with those obtained by XRD peak broadening profile analysis (Section \ref{sec5}).

\section{Experimental}
\subsection{Sample Preparation}
\label{sec2}

Commercially 1100 pure aluminum ($99.0\%$ pure) was used to perform RUS and XRD measurements. From the same as-received bar, five pieces were taken to prepare the studied conditions classified as original, annealed and laminated material: two samples were annealed at 673 K, one for 5 h and another for 10 h. Two others were cold-rolled, at either 33$\%$ or 43$\%$ of the initial diameter of the as-received original material.  It is well known that longer annealing leads to lower dislocation density, and stronger cold-rolling leads to higher dislocation density.  However, there is no known way to quantitatively estimate how much the dislocation density will change as a consequence of a given amount of annealing or cold-rolling. The pieces were thus numbered from 1 to 5, in the sense of expected increasing dislocation density. From each one of the five pieces, presumably with different dislocation densities, one portion was set aside for RUS testing, and another for XRD. 

The five RUS samples were shaped as rectangular parallelepipeds, with dimensions presented in 
Table \ref{tab:datosSamples}. 
Opposite sides are parallel within 0.06$^\circ$ and adjacent sides are orthogonal within 0.3$^\circ$. Our samples can be then modeled as perfect rectangular parallelepipeds \cite{Spoor,spoor96}. Samples are labeled $1,2,3,4$ and $5$ for increasing expected dislocation density. 

\begin{table}[h!]
\hspace{-2em}
\begin{tabular}{|c|c|c|c|c|c|} \hline \hline
&  Sample 1  &Sample 2 &Sample 3  & Sample 4  &Sample 5  \\ \hline
 & Annealed  & Annealed  & Original  & Rolled   & Rolled  \\
  & 673 K /10 h & 673 K/5 h &  & at 33\%  & at 43\%  \\ \hline

$d_1$ (cm)&$1.701 \pm 0.001$&$1.700 \pm 0.001$& $1.7011 \pm 0.0002$  & $1.696 \pm 0.001$  & $\,1.7007 \pm 0.0006\,$ \\

$d_2$ (cm) & $\,1.0015 \pm 0.0002\,$& $\,0.9997 \pm 0.0004\,$&  $\,0.9998 \pm 0.0003\,$  & $\,1.001 \pm 0.001\,$  &  $\,1.001 \pm 0.001\,$ \\

$d_3$ (cm)&$4.902 \pm 0.001$ &$4.900 \pm 0.001$ &  $4.901 \pm 0.001$  & $4.901 \pm 0.001$   & $4.900 \pm 0.001$  \\

$M $ (g)& $22.45\pm 0.01$ & $22.38\pm 0.01$ & $22.43\pm 0.01$ & $22.35\pm 0.01$ & $22.42\pm 0.01$ \\

$\rho$ (g/cm$^3$)& $2.688 \pm 0.002$ & $2.687 \pm 0.002$ &  $2.691 \pm 0.002$ & $\,2.687 \pm 0.004\,$ &  $2.687 \pm 0.003$ \\
\hline \hline
\end{tabular}
\caption{{\bf Sample characteristics:} Rectangular parallelepiped dimensions, mass and mass density for the five samples. Columns are ordered for increasing expected dislocation density.}
\label{tab:datosSamples}
\end{table}

All conditions tend to have anisotropic grain size distribution due to the conformation by extrusion in the case of original (as-received) and annealed bars, and due to lamination in the case of laminated conditions. Considering this, all samples for XRD were cut in the transversal axial direction from the original and annealed bars, and transversal to the laminated direction in laminated conditions. XRD samples were further chemically attacked with a solution of 30$\%$ HCl, 10$\%$ HF and 60$\%$ distilled water to remove superficial deformation induced during sample cut. A Siemens D5000 diffractometer was used for XRD measurements, using Cu K$\alpha$ (1.5418 \AA )radiation. The instrument-broadening contributions were measured independently using (111) and (100) oriented silicon single crystals, and were subtracted from the data prior to calculations \cite{balzar}. Si(111) is located at an angle of 28.426$^{\circ}$ while Si(400) is at 69.192$^{\circ}$, with a broadening measure at the Full Width at Half Maximum (FWHM) of 0.022$^{\circ}$ and 0.034$^{\circ}$, respectively.  Si(111) FWHM was subtracted in the case of the Al(111) and Al(200) signals while Si(400) was used for Al(311) and Al(222). Three XRD measurements were performed for samples 1, 2, 4 and 5, while two measurements were performed for sample 3.

\subsection{Resonant Ultrasound Spectroscopy}
\label{sec3}

An in-house built RUS apparatus \cite{caru,maite}, similar to those presented in \cite{migliori93,leisure97} was used to measure the speed of elastic waves of the five aluminum samples. This was done through the measurement of the resonant frequencies of the samples for stress-free boundary conditions. These frequencies, in turn, provide the elastic constants $C_{ij}$  (in Voigt notation) after the mass and linear dimensions of the samples have been independently measured. In order to satisfy this boundary condition, samples are placed in between an emitter and a receiver, held by two opposite corners, the receiver being held by a set of springs mounted on a linear air bearing. The weight of the receiver part of the setup is such that, at equilibrium with the spring force, the distance between the emitter and receiver surfaces is slightly larger than the sample's diagonal. In this way, the sample can only be held by its corners in the setup if a small mass is added to the receiver part, typically of $10$ g. Thus, the sample-apparatus contact force is small, with an upper bound of  $0.1$ N, corresponding to about one half of the sample weight $Mg \approx 0.22$ N. The estimated contact area is  $\sim 0.1 - 0.2$ mm$^2$. 

\begin{figure*}[h!]
{\includegraphics[width=16cm]{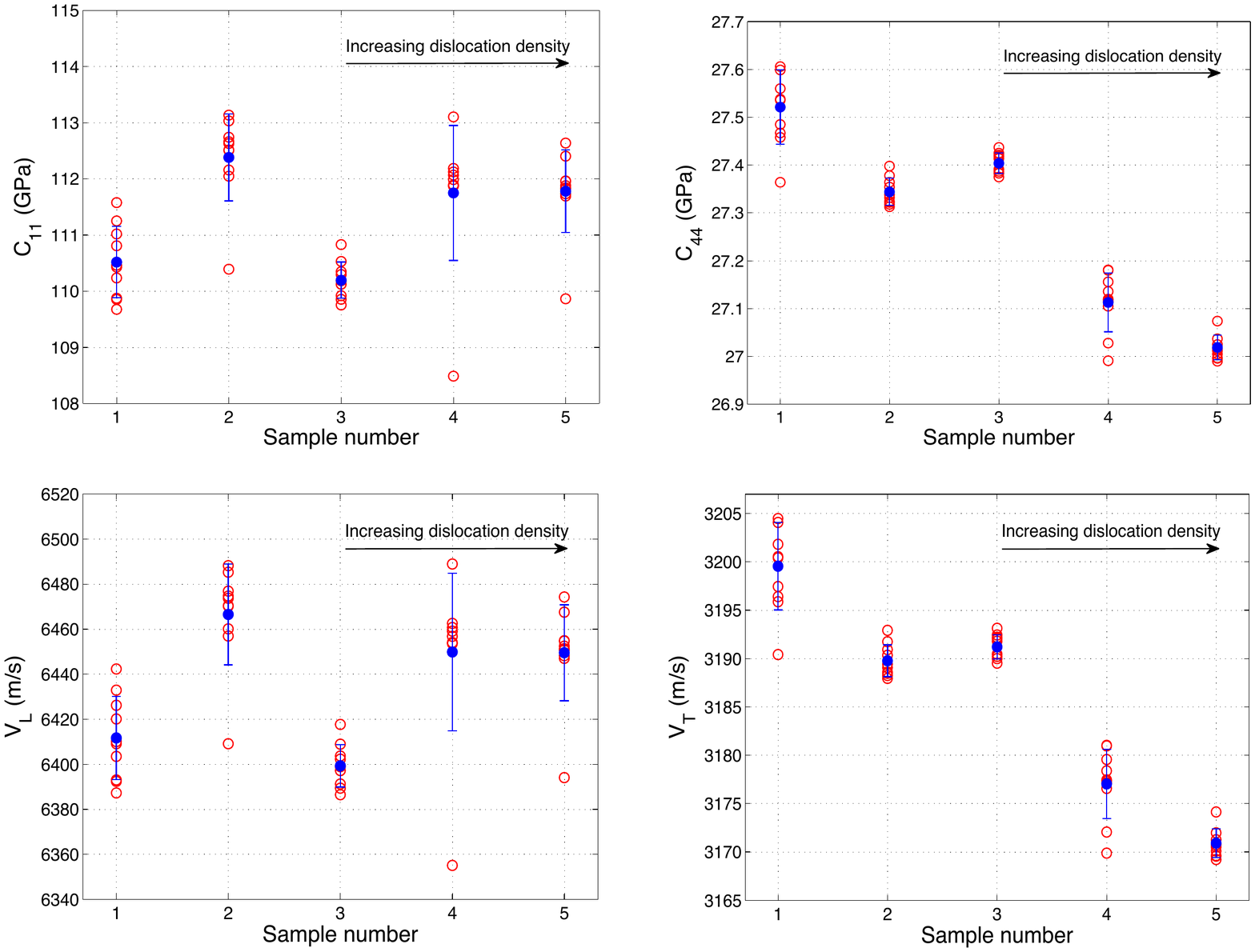}}
\caption{{\bf Elastic constants and wave velocities of samples with different dislocation content as measured by RUS.} Elastic constants $C_{11}$ and $C_{44}$ (top) and longitudinal ($v_L$) and transverse ($v_T$) (bottom) sound speeds as determined by RUS for the different aluminum samples. Open circles correspond to results obtained for ten different sample positionings and solid circles correspond to the mean values. Error bars are computed from standard deviations of the set of ten measurements. Note the difference in scale between the left-hand-side-panels and the right-hand-side ones. The shear modulus, $C_{44}$, is determined with much higher accuracy than $C_{11}$, allowing the speed of shear waves, $v_T$, to be determined with an accuracy of $\sim 0.1\%$. It exhibits a clear downward trend in the direction of expected increasing dislocation density.}
\label{Fig_Cjj}
\end{figure*}

The results of RUS measurements are shown in Table \ref{tab:datosRUS} and Figure \ref{Fig_Cjj}. Preliminary results, with a smaller data base, were reported in \cite{rus1}. A satisfactory fit to the data is obtained under the assumption of transverse isotropy, so that, of the 21 possible elastic constants, only five are independent. $C_{44}$ can be determined with much higher accuracy than the other elastic constants, a well known characteristic of RUS \cite{rusbook}. From it, and the independently measured mass densities, the speed of shear waves for each sample can be determined with an accuracy of $\sim 0.1\%$. 

\begin{table*}[h!]
\hspace{-2em}
\begin{tabular}{|c|c|c|c|c|c|c|} \hline \hline
& &  Sample 1  &Sample 2 &Sample 3  & Sample 4  &Sample 5  \\ \hline
& & Annealed  & Annealed  & Original  & Rolled   & Rolled  \\
&  &  673 K /10 h & 673 K /5 h &  & at 33\%  & at 43\%  \\ \hline

RUS & $C_{11}$ (GPa)& $111 \pm 1 $& $ 112\pm 1$  &  $110.2 \pm 0.3 $ & $112 \pm 1$ &  $112 \pm 1$   \\

& $C_{33}$ (GPa)& $112 \pm 1 $& $ 113\pm 1$  &  $111 \pm 1 $ & $113 \pm 3$ &  $112 \pm 1$   \\

& $C_{23}$ (GPa)& $61 \pm 1 $& $ 62\pm 1$  &  $60.5 \pm 0.4 $ & $63 \pm 2$ &  $61 \pm 1$   \\

& $C_{12}$ (GPa)& $58 \pm 1 $& $ 60\pm 1$  &  $58.2 \pm 0.3 $ & $60 \pm 2$ &  $60 \pm 1$   \\

& $C_{44}$ (GPa)&$\,27.52\pm 0.08\,$ & $\,27.34\pm 0.03\,$ &  $\,27.40 \pm 0.02 \,$ &  $\,27.11 \pm 0.06\,$ &  $\,27.02\pm 0.03\,$\\

& $v_L$ (m s$^{-1}$)&$6412\pm 19$ & $6467\pm 22$ &  $6399 \pm 9 $ &  $6450 \pm 35$ &  $6450\pm 21$\\

& $v_T$ (m s$^{-1}$)&$3200\pm 5$ & $3190\pm 2$ &  $3191 \pm 1 $ &  $3177 \pm 4$ &  $3171\pm 2$\\

XRD & $\Lambda (10^8$ mm$^{-2}$)&$11.4\pm0.5$ & $9.1\pm4.5$ &  $14.3\pm8.9$ &  $63\pm8$ &  $65\pm6$\\

& $D$ (nm) & $86 \pm 27$  & $82 \pm 17$ & $94 \pm 22 $ & $130 \pm 34$ & $120 \pm 36$  \\

\hline \hline
\end{tabular}
\caption{{\bf RUS and XRD measurements results.} Top seven lines: Elastic constants ($C_{11}$, $C_{33}$, $C_{23}$, $C_{12}$ and $C_{44}$ in Voigt notation), longitudinal wave velocity ($v_L$), and shear wave velocity ($v_T$) obtained through RUS. Columns are ordered for increasing expected dislocation density. Absolute errors for $C_{ij}$ constants are computed from the standard deviations of ten RUS measurements obtained for each sample. The errors of $v_L$ and $v_T$ are obtained from the standard deviations of the set of ten values obtained from their definitions $v_L = \sqrt{C_{11}/\rho}$ and $v_T= \sqrt{C_{44}/\rho}$, with $\rho$ the mass density. For transverse isotropy, the other elastic constants are $C_{22} = C_{11}$, $C_{13} = C_{23}$, $C_{55} = C_{44}$ and $2C_{66} = C_{11}-C_{12}$. Note that $C_{44}$ can be determined with much better accuracy than the other constants. Bottom two lines:  Dislocation density $\Lambda$ and crystallite size $D$, obtained by XRD. Absolute errors for $\Lambda$ and $D$ are computed from the standard deviations of three XRD measurements performed for samples 1, 2, 3 and 4, and for two measurements performed for sample 3.}
\label{tab:datosRUS}
\end{table*}

For each one of our five samples a set of 30 resonant frequencies was measured, starting from the lowest resonant mode around $19$ kHz and finishing at $\sim150$ kHz. The pure elastic, non-dissipative assumption is well verified: the lowest quality factor $Q$ is $\sim 10^3$. In order to verify that all resonance frequencies are measured in the elastic linear regime, each resonant curve is measured for five different driving amplitudes. We then verify that the resonance amplitude is linear with the driving amplitude. We emphasize that the main difference with our  previous, preliminary, results presented in \cite{rus1} is the number of independent measurements for each sample. For the results presented in the present manuscript each sample was placed ten times in the apparatus in order to reduce errors due to slight dependence of the resonant frequencies on the contact load and positioning with respect to the ultrasonic receiver. In \cite{rus1}, each sample spectrum was measured just once. In fact, there is an incorrect phrase in \cite{rus1} that states that each sample was measured five times. This was the case in early measurements for a lower number of measured frequencies ($12$) assuming complete isotropy. 

As briefly reported in \cite{rus1}, attempts to fit resonant frequencies assuming homogeneity and isotropy fail considerably. In fact, under the assumption of isotropy, measurements of $C_{44}$ agree with previous published results but not those of $C_{11}$, which are about $20\%$ lower than the values reported in the literature. This discrepancy arises independent of sample preparation, positioning and the particular emitting transducer type. The complete isotropy hypothesis is then questionable. Indeed, our aluminum samples are prepared with their longest dimension in the extrusion direction of the original aluminum bar and microscopic images (not shown) confirm that crystal grains are elongated in this direction.  We then consider that our samples are transversely isotropic, which implies that the elastic-constant matrix has five independent values. Having that in consideration the measured resonant frequencies fit much better the theoretical predictions. It is important to stress that this is not just due to the increment in the number of adjustable parameters, it is truly a consequence of a better assumption concerning the material structure. To prove this, we have performed anisotropy tests by varying the anisotropic and isotropic planes in the RUS inverse code.  The better results, quantified by the final root-mean-square error between measured and predicted frequencies, correspond to those where the sample's longest dimension is the anisotropic one with respect to the transverse isotropic dimensions. The measured elastic constants, as well as the longitudinal ($v_L$) and shear ($v_T$) wave velocities related to $C_{11}$ and $C_{44}$ respectively, are given in Table I. Figure 1 shows $C_{11}$, $C_{44}$, $v_L$ and $v_T$ versus sample number. Each elastic constant is obtained by averaging 10 results obtained from 10 independent sample positionings. The reported errors correspond to standard deviations of these 10 measurements. As expected, deviations from isotropy are small, for example $C_{11}/C_{33} >0.99$ and $C_{44}/C_{66} < 1.06$. Thus, a comparison with the theory presented below is possible. No clear tendency can be observed for $v_L$ because of the large experimental errors, i.e., $15\%$. However, $v_T$ data is much more precise,  errors $\approx 0.03\% - 0.15\%$, and a clear decreasing tendency in the direction of expected increasing dislocation density is observed.

\section{Interpretation of RUS measurements}
\label{sec4}

In order to link the RUS measurement results to dislocation density we use the formula \cite{ijbc1,Maurel5}(see Appendix \ref{app1})
\beq
\frac{\Delta v_T}{v_T} =
-\frac{8}{5\pi^4} \frac{\mu b_s^2}{\Gamma_s} \Delta( n_s L_s^3) -
\frac{4}{5\pi^4} \frac{\mu b_e^2}{\Gamma_e} \Delta( n_e L_e^3)
\label{muR} \, .
\eeq
Eqn. (\ref{muR}) links the relative change in shear wave velocity ${\Delta v_T}/{v_T}$ between two samples of a material  that differ in dislocation density $n$, where $n$ is the number of dislocation segments of length $L$, Burgers vector $\vec b$ and line tension $\Gamma$, per unit volume. The subscript $e$ is for edge dislocations, $s$ is for screws, and $\mu$ is the shear modulus of the reference material. This is a result valid to leading order in perturbation theory when the changes in $v_T$ are small.  Formula (\ref{muR}) shows that an increasing dislocation density results in a decreasing speed of shear waves. It is obtained using Multiple Scattering Theory, a reasoning that quantifies the intuition that dislocations will make it more difficult for waves to propagate \cite{ishimaru}. The physical process that is responsible for the change in the speed of shear waves is as follows: an elastic wave is incident upon a dislocation segment of length $L$ whose ends are pinned. As a result, it oscillates like a vibrating string. This oscillation, in turn, generates secondary waves. As explained in Appendix \ref{app1}, the coherent superposition of many such scattering processes leads to Eqn. (\ref{muR}). 

Considering that \cite{Lund88}
\bea
\Gamma_e & = & \mu b^2_e \left( 1-\frac{v_T}{v_L} \right) \\ 
\Gamma_s & = & \mu b^2 _s 
\label{eqlinetension}
\eea
where $v_L$ is the speed of longitudinal waves, we have
\beq
\frac{\Delta v_T}{v_T}  \approx  - \frac{8}{5\pi^4} \left(   \Delta ( n_s L_s^3) +  \Delta ( n_e L_e^3 ) \right)  \equiv - \frac{8}{5 \pi^4}  \Delta ( nL^3) \, .
\label{eqdeltav} 
\eeq
What RUS does, then, is to provide a measure of the dimensionless quantity $nL^3$. This formula does not provide an absolute measurement of dislocation density, but a measurement of the difference in dislocation density between two samples. The data show that our RUS measurements cannot, within experimental error, differentiate between samples 2 and 3, but can differentiate sample 1 from sample 2, sample 3 from sample 4, and sample 4 from sample 5, providing the values of dislocation densities given in Table \ref{table3}. We have $(\Delta v_T / v_T) \sim  \Delta (nL^3)/60$ with the values given in Table \ref{table3}. Of course, the same value of $\Delta ( nL^3)$ can be obtained with a variety of values for $n$ (number of segments per unit volume) and $L$ (distance between pinning points) separately. Taking, for visualization purposes $L \sim 10$ nm, we get that the samples differ in dislocation density $\Delta  \Lambda \sim (1-3) \times 10^9$ mm$^{-2}$ (in the usual units).

\begin{table*}[h!]
\hspace{4em}
\begin{tabular}{|c|c|c|c|c|} \hline \hline
&  Samples 1-2  &Samples 2-3 &Samples 3-4  & Samples 4-5   \\  \hline

$\frac{\Delta v_T}{\langle v_T \rangle} \times 10^3$ & 3.1$\pm$ 1.7 & - 0.3 $\pm$ 0.7 &  4.4 $\pm$ 1.3  &  1.9 $\pm$ 1.4 \\

$\Delta (nL^3)$ & 0.19 $\pm$ 0.11 & - 0.02 $\pm$ 0.04 & 0.28 $\pm$ 0.08 & 0.12 $\pm$ 0.09   \\

\hline \hline
\end{tabular}
\caption{{\bf RUS measurement of dislocation density.} RUS measures the difference $\Delta v_T$ in the speed of shear waves between the various samples. Results are given in the first line relative to the arithmetic mean $\langle v_T \rangle$ of each pair of values. From these values and using Eqn. (\ref{eqdeltav}), the difference in dislocation density $\Delta (nL^3)$ can be obtained (second line). Within experimental error, it is not possible to distinguish the difference in dislocation density between  samples 2 and 3.}
\label{table3}
\end{table*}

We now provide validation to the dislocation density values obtained through RUS with XRD peak-broadening measurements on the same samples.

\section{X-ray diffraction analysis}
\label{sec5}
 
A typical XRD pattern obtained from the aluminum samples is shown in Figure 2. The characteristic (111), (200), (311) and (222) signals are observed while the (220) reflection is absent. The absence of the (220) reflection is due to texturing induced by the deformation of the sample, a well known phenomenon in metallurgy \cite{metal}. For all peaks present in the diffractograms we observed that the larger the matrix deformation (from the annealed to the extreme laminated conditions) the larger the peak width. The inset in figure 2 shows this effect for the Al(111) reflection. 

 \begin{figure}[h!]
 \hspace{8em}
\includegraphics[width=9cm]{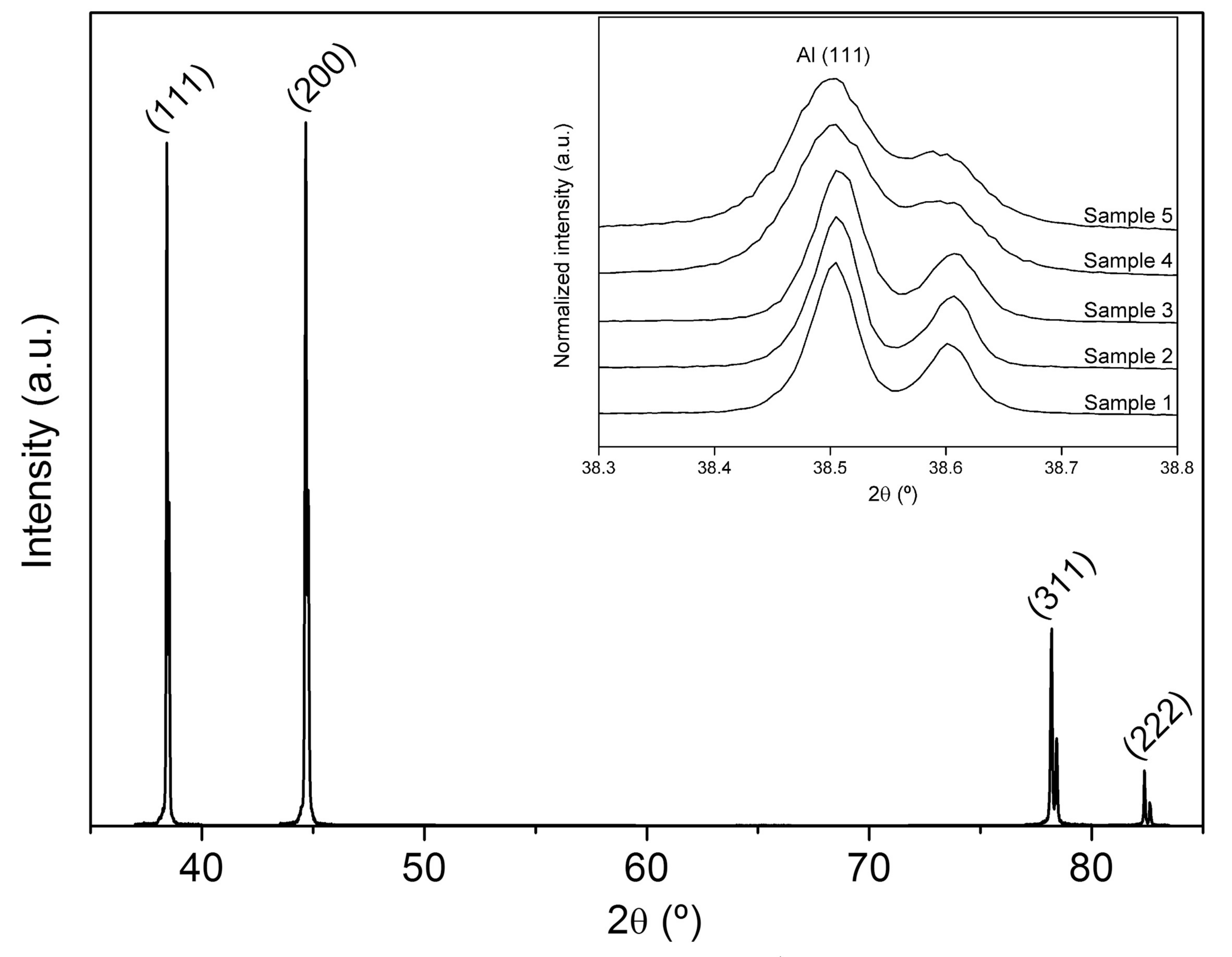}
\caption{{\bf XRD pattern for Sample 2.} The characteristic (111), (200), (311) and (222) signals are observed while the (220) reflection is absent due to texturing induced by the deformation of the sample. The inset shows a zoom of the (111) peak of the XRD patterns for the five samples.}
\label{Fig_XRD1}
\end{figure}

Current theory \cite{ungar} accounts for the broadening of
XRD peaks as arising from two effects, finite crystallite size and
presence of dislocations. If the two effects are uncorrelated, they will act additively:
\beq
\Delta K = {\Delta K}^S + {\Delta K}^D
\eeq
where $\Delta K$ is the Full Width at Half Maximum (FWHM) of the diffraction peak at wave vector $K$. ${\Delta K}^S$ is the contribution of the crystallites finite size, and ${\Delta K}^D$ is the contribution of the dislocations.  On dimensional grounds ${\Delta K}^S = C_1/D$ where $D$ (dimensions length) is the crystallite size, and $C_1$ is a dimensionless constant. This quantity is independent of the X-ray wavelength.

Let us estimate ${\Delta K}^D$ on dimensional grounds \cite{barenb}. It must depend on $b = |\vec b|$, the magnitude of the Burgers vector of the dislocations (assuming only one value for its magnitude and taking an average if there are several; in any case its value is of the order of the size of the cube root of the volume of the unit cell). Also, it must depend on $\Lambda$, the dislocation density (dimensions length$^{-2}$), assumed uniform. Consequently, it must involve the product $b\Lambda^{1/2}$, since the broadening must be an increasing function of $b$ (larger $b$ means greater deformation, hence larger broadening). For low dislocation densities (i.e. such that $\Delta K$ is small compared to $K$), its dependence must be linear. There is an additional quantity with dimensions of inverse length,  $K$, the diffraction vector of the broadened peak, and for dimensional analysis consistency we get 
\beq
{\Delta K}^D  =   C_2^{1/2} \, b \, \Lambda^{1/2} \, K   \, .  \\
\label{deltakd}
\eeq 
There cannot be an added contribution from the dimensionless quantity $bK$ since in that case there would be a broadening even for vanishing dislocation density $\Lambda$. Here $C_2$ is a dimensionless constant that
may depend on the crystalline plane (Miller indices ($h,k,l$))
doing the scattering, but it does not depend on
$\Lambda$. Higher powers of $b$ (with concomitant higher powers
of $K$) seem unlikely, again claiming linearity at the lowest
order. This type of functionality is deduced through the modified Williamson-Hall method \cite{re9996,re9999} from an analysis of the full shape
of the spectral lines (see Appendix \ref{sec_xrd_theo}), with the result that
\beq
\left ( \Delta K \right )^2 =  \left ( \frac{0.9}{D}\right )^2   + \left ( \frac{\pi M^2 b^2 \Lambda}{2}\right ) \bar{C}_{h00} (1-qH^2) K^2 
\label{EcnDK2}
\eeq
where $\bar{C}_{h00}$ and $q$ can be expressed in terms of the elastic constants of the crystal and empirically determined parameters, $H^2$ is the fourth order invariant of the $hkl$ indices of the different reflections and $M^2$ is a constant depending on the effective outer cut-off radius of dislocations. We take $M^2 = 0.1$ \cite{re9996,re9999}. Consequently, a plot of  $\left ( \Delta K \right )^2 $ vs $K^2$ will yield a straight line: The  slope determines the dislocation density $\Lambda$, and the intercept with the vertical determines the crystallite size $D$. Eqn. (\ref{EcnDK2}) deals with $( \Delta K )^2$, the square of the FWHM. The result is equivalent to the result for $\Delta K$ obtained in (\ref{deltakd}) using dimensional analysis as long as crystallite size and dislocation density are uncorrelated \cite{re13}.

\subsection{Modified Williamson-Hall plot}

\begin{figure}[h!]
 \hspace{8em}
{\includegraphics[width=9cm]{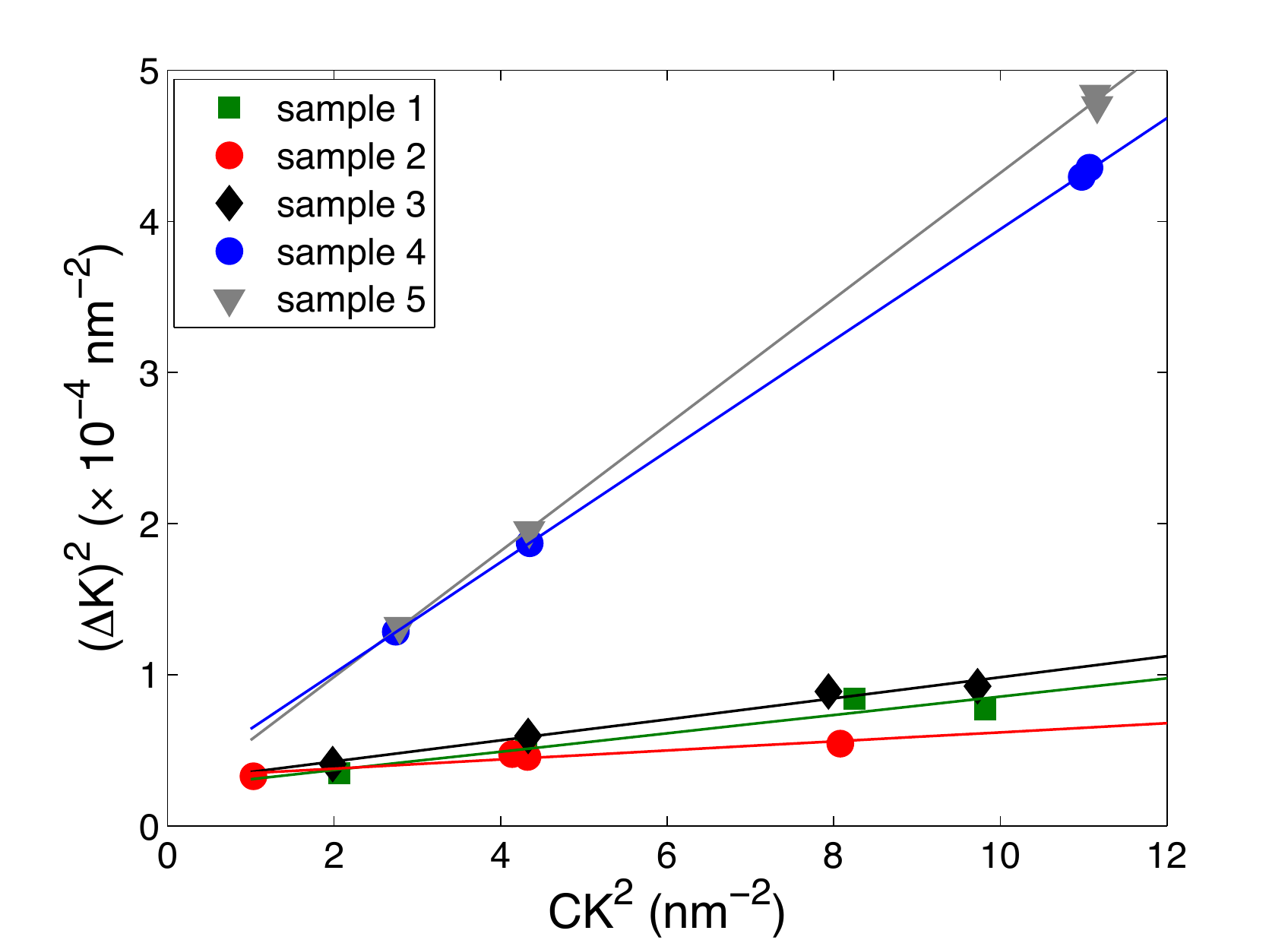}}
\caption{{\bf $\Delta K^2$ versus $K^2 C$ plot for the five different aluminum samples, obtained through the modified Williamson-Hall method.} According to Eqn. (\ref{EcnDK2}) the slope provides a measure of the dislocation density, and the intercept with the vertical axis provides a measure of the crystallite size. The resulting values are quoted in Table \ref{tab:datosRUS}.}
\label{Fig_XRD2}
\end{figure}

Figure \ref{Fig_XRD2} shows the $(\Delta K)^2$ vs. $K^2 C$ plot calculated according to the `modified' Williamson-Hall method from the values obtained of the diffraction patterns following Eqn. (\ref{EcnDK2}). The increasing slope from annealed to laminated conditions indicates an increasing dislocation density according to Eqn. (\ref{EcnDK2}). The values of dislocation density $\Lambda$ for all conditions are shown in Table \ref{tab:datosRUS}. The intercepts with the vertical axis are, within experimental error, similar, indicating no significant change in the crystallite size $D$ among the different samples, so that the X-ray peak broadening is due only to the dislocations (Table \ref{tab:datosRUS}).

\section{Conclusion}
Qualitatively, Figure \ref{Fig_Cjj} shows that, as expected, the shear modulus $C_{44}$ of the different samples decreases in the direction of expected dislocation density. Also, Figure \ref{Fig_XRD2} shows that XRD lines also broaden, as expected, in the direction of expected dislocation density. Figures \ref{Fig_Cjj} and \ref{Fig_XRD2} show that the five samples can be separated in two groups: Samples 1, 2 and 3 with low dislocation density, and  samples 4 and 5 with clearly large dislocation density. Interestingly, ultrasound and XRD measurements coincide in their diagnostic of differentiating the two groups and which samples have the larger number of dislocations.

\begin{figure}[h!]
\hspace{8em}
{\includegraphics[width=9cm]{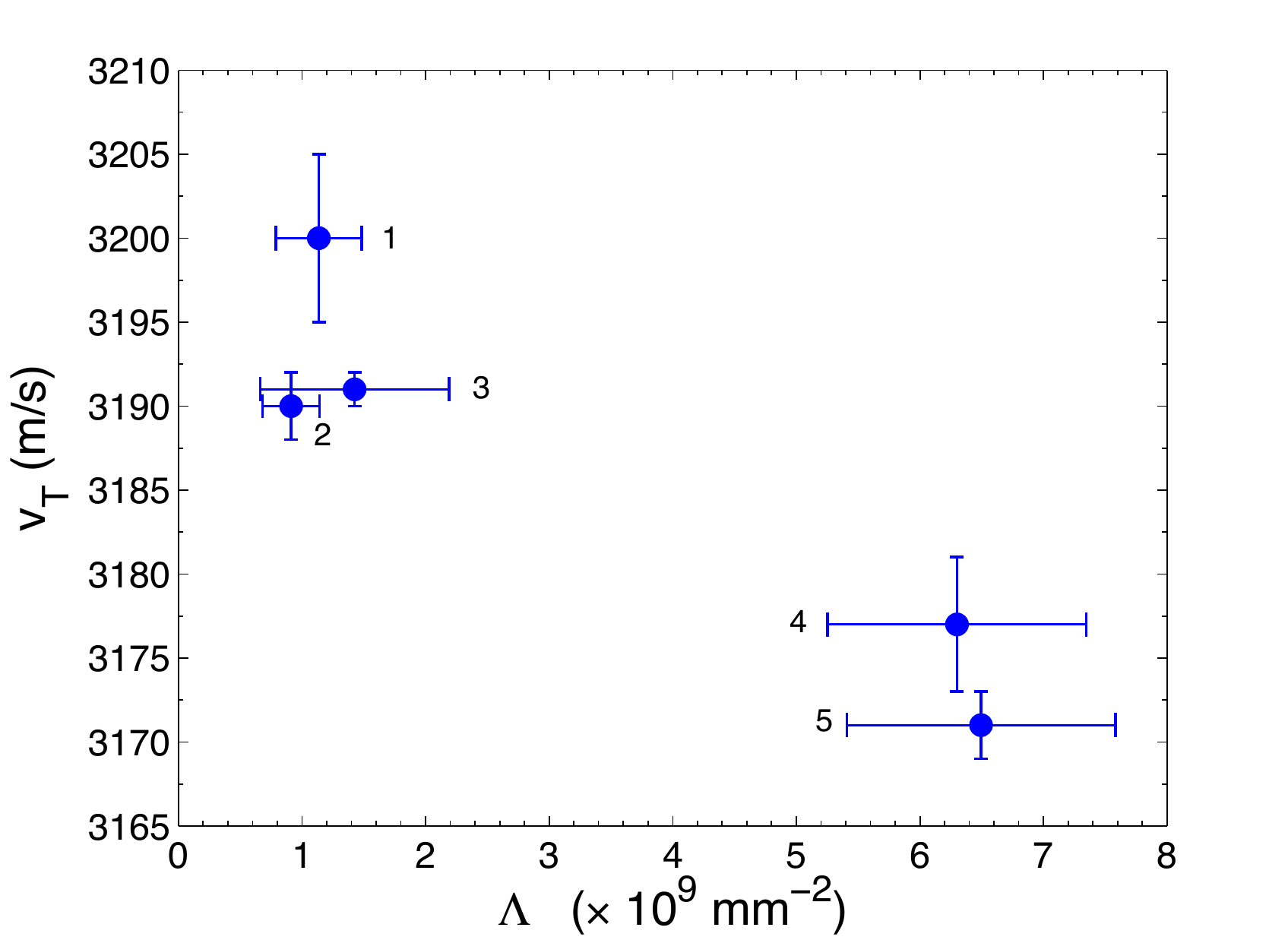}}
\caption{{\bf RUS and XRD characterization, compared and contrasted.} Shear wave velocity $v_T$ determined through RUS, plotted against the dislocation density $\Lambda$ determined through XRD, for the five aluminum samples. RUS coincides with XRD in identifying two groups: samples 1, 2 3, and samples 4, 5. However, RUS goes beyond XRD since it can differentiate 1 from 2 and 3, 2 from 4, and 4 from 5. The resulting values for $\Delta (n L^3)$ are quoted in Table \ref{table3}. }
\label{Fig_vt_rho}
\end{figure}

Figure \ref{Fig_vt_rho} provides a comparison between the results obtained with RUS and those obtained with XRD. Quantitatively, RUS is a better discriminant of dislocation density among these five aluminum samples than XRD: The latter can distinguish, within experimental error, two different values, while the former can distinguish four.  Now, ``dislocation density'' has a slightly different meaning in the two contexts: In XRD, the measurements are modelled in terms of $\Lambda$, with dimensions inverse length squared, defined as the total length of dislocation line within a sample divided by sample volume. In RUS, measurements are modelled in terms of $nL^3$, with $L$ the length of the vibrating dislocation line segments, or distance between dislocation pinning points, and $n$ the number of such segments per unit volume. Although the theory of \cite{Maurel5} is sufficiently general to accomodate dislocation segments of different length, here we take all segments of the same, average, length for simplicity. The two values of $\Lambda$ detected by XRD are $\sim 10^9$ mm$^{-2}$ (samples 1, 2 and 3) and $\sim 6 \times 10^9$ mm$^{-2}$ (samples 4  and 5).  RUS does not provide absolute values of $nL^3$, but  differences between different samples. The precise numbers depend on the nature (screw, edge, prismatic) of the dislocations.  In our case we can discriminate values of $\Delta(nL^3)$ in the range $0.1-0.3$. If we take, tentatively, $\Lambda = nL$ and the values of $\Lambda$ provided by XRD, we can conclude that samples 1 and 2 have the same $\Lambda$ but different $L$, and the same for the samples 4 and 5. But samples 3 and 4 differ both in $n$ and $L$. 

Finally, our XRD measurements show a texture effect due to the different samples history, that could lead to missing part of the dislocation density information. Eventually, this could be avoided taking cuts in the longitudinal direction of the material for the different conditions \cite{re16}. In this sense, RUS presents an advantage over XRD because it takes an average over the whole sample in every measurement. 

In conclusion, our results clearly establish that RUS can distinguish between aluminum samples with different dislocation densities, thus opening the way for the use of ultrasonic techniques in the characterization of the plastic behavior of materials.

\section*{Acknowledgements} Useful discussions with F. Barra, A. Sep\'ulveda and R. Schwarz are gratefully acknowledged. This work was supported by Fondecyt Grant 1100198 and Anillo Grant ACT 127.

\appendix

\section{Interaction of elastic waves with dislocations} 
\label{app1} 

The theory that we use to interpret the RUS measurements is based on recent work by Maurel et al. in which the scattering of an elastic wave of arbitrary polarization by a pinned (edge) dislocation segment \cite{Maurel4}  of finite length $L$, or an individual prismatic dislocation loop \cite{natalia} was calculated in detail. The ensueing results were used within the framework of multiple scattering theory to develop formulae for the effective wave velocity, and attenuation, of an elastic wave propagating within an elastic medium filled with edge dislocation segments, or prismatic dislocation loops \cite{natalia}. The case of screw dislocations is calculated below. These developments build on earlier results of  Granato and L\"ucke \cite{gl1,gl2} and their primary advantage for the purpose of the present work is the ability, absent in \cite{gl1,gl2}, to distinguish between longitudinal and transverse waves. This ability is essential  to interpret the RUS results.

Consider an homogeneous, isotropic, elastic continuum, of density $\rho$, shear modulus $\mu$ and bulk modulus $\cal B$. This is the reference material. It does not matter whether it does or does not have dislocations to begin with. All that matters is that it can be described as homogeneous, continuous and elastic.  Next, take the same medium with an additional random (uniform) distribution of dislocations: density (number per unit volume) $n_e$ (resp. $n_s$) of pinned edge (resp. screw) dislocation segments of length $L_e$ (resp. $L_s$) and Burgers vector $b_e$ (resp. $b_s$). If the material is loaded by a monochromatic elastic wave, the dislocations will interact with the wave and will influence the material's response. The model used in all cases is the string model of Koehler \cite{koehler}: The dislocation is treated as an infinitely thin string endowed with mass and elastic tension; as it is loaded by the elastic wave it responds by oscillating along glide planes and, as it oscillates, it re-radiates. This process leads to a new (``renormalized'') value $\mu^R$ for the shear modulus. The bulk modulus does not change. The result is, in the limit of low frequencies,
\beq
\frac{\mu^R}{\mu} -1 =
-\frac{16}{5\pi^4} \frac{\mu b_s^2}{\Gamma} n_s L_s^3 -
\frac{8}{5\pi^4} \frac{\mu b_e^2}{\Gamma} n_e L_e^3 
\label{muRmu} \, ,
\eeq
from which Eqn. (\ref{muR}) follows. An equation, qualitatively similar to (\ref{muRmu}) that links dislocation density with shear modulus decrement, has been provided in \cite{nl1}.  The formulae of \cite{nl1} do provide the correct scaling of said decrement in terms of dislocation length. However, they depend on undetermined multiplicative parameters, the resolving shear factor and the conversion factor from shear strain to longitudinal strain. The reasoning below provides a justification of the numerical coefficients presented in (\ref{muRmu}). It also provides a justification of the result of \cite{nl1} as the low-frequency limit of a finite-frequency formulation.

We shall ignore the effect of prismatic dislocation loops. We now derive the result (\ref{muRmu}) for screw dislocations. The reasoning follows closely an analogous calculation of \cite{rus1} for edge dislocations.

Elastic waves in an homogeneous, isotropic, three-dimensional, infinite elastic medium are described by displacements $\vec u (\vec x, t)$ as a function of equilibrium position $\vec x$ and time $t$ that satisfy the wave equation
\beq
\rho \frac{\partial^2 u_i}{\partial t^2}  -
c_{ijkl}\frac{\partial^2 u_k}{\partial x_j \partial x_l} = 0
\label{eq1}
\eeq
with $c_{ijkl} = \lambda \delta_{ij} \delta_{kl} + \mu (\delta_{ik} \delta_{jl} + \delta_{il} \delta_{jk})$ the tensor of elastic constants, and $i,j,k, = 1,2,3$ which in the isotropic case has only two independent constants: $\mu$, and $\lambda = [{\cal B} - (2\mu) /3]$. Such a solid supports two types of propagating waves: longitudinal (acoustic) and transverse (shear) waves with propagation velocity  $v_L = \sqrt{(\lambda +2\mu)/\rho}$ and $v_T = \sqrt{\mu/\rho}$, respectively. Their ratio $\gamma = v_L/v_T$ is always greater than one, i.e. $\gamma > 1$.

\vspace{0.5em}
Dislocations are modeled as ``strings'' of length $L$, mathematically described through a position vector $\vec X (s,t)$, with $s$  a Lagrangean parameter to label points along the string,  which is pinned at the ends. Their simplest equilibrium position is a straight line. They are characterized by a Burgers vector $\vec b$, perpendicular to the equilibrium line for an edge and parallel to it for a screw. Their unforced motion is described by a conventional vibrating string equation
\beq
m\frac{\partial^2 X_i}{\partial t^2} + B \frac{\partial X_i}{\partial t}
-\Gamma \frac{\partial^2 X_i}{\partial s^2} = 0
\label{eq2}
\eeq
where the mass per unit length $m$ and line tension $\Gamma$ for screw dislocations are given by \cite{Lund88} 
\bea
m_s & = & \frac{\rho b^2}{4\pi}  \ln \left( \frac{\delta}{\delta_0} \right) \nonumber \\
\Gamma_s & = & \frac{\mu b^2c_T^2}{4\pi}  \ln \left( \frac{\delta}{\delta_0} \right) \, .
\label{eqml}
\eea
$\delta$ and $\delta_0$ are external and internal cut-offs, and $B$ is phenomenological drag coefficient.  We shall only consider glide motion since climb involves diffusion, which happens, at room temperature, on a time scale long compared to the inverse of the wave frequencies under consideration.

\vspace{0.5em}
Eqns. (\ref{eq1}) and (\ref{eq2}) describe waves and dislocations that do not interact. The interaction is introduced through right hand side---source---terms. Also, it is convenient to describe the waves not through particle displacement $\vec u$ but in terms of particle velocity $\vec v = \partial \vec u / \partial t$. When $N$ dislocations are present this leads to \beq
\rho \frac{\partial^2 v_i}{\partial t^2}  -
c_{ijkl}\frac{\partial^2 v_k}{\partial x_j \partial x_l} = s_i
\label{eq1s}
\eeq
where the right-hand-side term $s_i$ is given by
\bea
 s_i(\vec x,t) & = & c_{ijkl}\epsilon_{mnk} \sum_{n=1}^N\int_{\cal L}
d s \; \dot X_m^n( s ,t) \tau_n b_l  \nonumber \\
 & & \mbox{} \times \frac{\partial}{\partial x_j}
\delta ( \vec x-\vec X^n (s,t) ).
\label{termesourceN}
\eea
Here, $\epsilon_{mnk}$ is the completely antisymmetric tensor of order three, $\hat{\tau}$ is a unit tangent along the dislocation line. For screw dislocations in equilibrium, the Burgers vector points along this tangent: $\vec b = b \hat{\tau}$. In the case of the string equation (\ref{eq2}) the coupling with elastic waves is provided by  the Peach-Koehler \cite{pk} force :
\begin{equation}
m {\ddot{X}}_k( s ,t)+ B{\dot{X}}_k( s ,t)- \Gamma X_k''( s ,t)= \mu b \; {\mathsf
  N}_{kjp}\partial_j u_p( \vec X ,t) ,
\label{eqmouv2}
\end{equation}
with ${\mathsf N}_{kjp}\equiv \epsilon_{kjm} \tau_m \tau_p +
\epsilon_{kpm} \tau_m \tau_j$. Overdots mean time derivatives, and primes mean derivatives with respect to $s$.

\vspace{0.5em}
The procedure now is as follows: The loaded string equation (\ref{eqmouv2}) is solved in terms of normal modes, and the solution plugged into the right hand side of the wave equation (\ref{eq1s}). In the long wavelength limit, $\lambda \gg L$ (for $L$ up to 100 nm this allows for frequencies well into the hundreds of MHz regime), and for small string displacements, the result of this operation is
\beq
-\rho \omega^2 v_i   -
c_{ijkl}\frac{\partial^2 v_k}{\partial x_j \partial x_l} = V_{ik} v_k
\label{eqmany}
\eeq
where 
\begin{equation}
V_{ik}=  \frac{8 L}{\pi^2}\frac{(\mu b)^2}{m}
\frac{S(\omega)}{\omega^2}  \; \sum_{n=1}^N {\mathsf N}^n_{mij}
 \frac{\partial}{\partial x_j}  \delta (\vec x- \vec X^n_0 ) \;
{\mathsf  N}^n_{mlk}{\frac{\partial}{\partial x_l}}
\label{potentialik}
\end{equation}
with
$S(\omega) \simeq
  {\omega^2}/(\omega^2-\omega_1^2 + i\omega B/m)$
and
$\omega_1$ is the frequency of the fundamental mode of the string with fixed ends: $\omega_1 = ({\pi}/{L}) \sqrt{\Gamma/m}$.

\vspace{0.5em}
The next step is to consider wavelengths long compared to the mean distance between dislocations, an assumption that is well satisfied by our experimental conditions involving centimeter-sized samples, so that the discrete sum in (\ref{potentialik}) is smoothed, replacing it with an integral over space with a continuous density $n(\vec x)$ of dislocation segments, and the tensor ${\mathsf N}^n_{mij} {\mathsf  N}^n_{mlk}$ by its angular average, $\langle {\mathsf N}^n_{mij} {\mathsf  N}^n_{mlk} \rangle$, assuming all directions equally likely. A reasoning like this one is used to study waves in plasmas \cite{Stix1992} and it is valid for wavelengths long compared to inter-dislocation distance.  It is straightforward to check that
\beq
\langle {\mathsf N}^n_{mij} {\mathsf  N}^n_{mlk} \rangle = \frac 25 \left( \delta_{il} \delta_{jk} + \delta_{ik} \delta_{jl} \right) - \frac{4}{15} \delta_{ij} \delta_{lk}
\label{av}
\eeq
Eqn. (\ref{eqmany}) thus becomes, in the case of uniform dislocation density $n(\vec x) = n$,
\bea
\lefteqn{-\rho \omega^2 v_i  -
c_{ijkl}\frac{\partial^2 v_k}{\partial x_j \partial x_l} =}  \nonumber \\
 & & {\cal A} \left[-\frac{4}{15} \delta_{ij} \delta_{kl} + \frac 25 (\delta_{ik} \delta_{jl} +\delta_{il} \delta_{jk})   \right] \frac{\partial^2 v_k}{\partial x_j \partial x_l}
\eea
where
\[
{\cal A} = (b\mu)^2 \frac{8}{\pi^2} \frac Lm \frac{1}{\omega^2 - \omega_1^2 + i\omega B/m} n
\]
From this expression it is easy to read the renormalized values of the Lame constants:
\bea
\lambda^R & = & \lambda - \frac{4}{15} \cal A \nonumber \\
\mu^R & = & \mu + \frac 25 \cal A
\eea
We see that $3 \lambda^R +2 \mu^R = 3 \lambda +2 \mu$. Finally, for low frequencies, so $\omega \ll \omega_1$ and $\omega \ll B/m$, we get the first term on the right-hand-side of Eqn. (\ref{muRmu}).

\section{XRD theory: Peak broadening due to the presence of dislocations}
\label{sec_xrd_theo}

Using quantitative models it is possible to correlate microstructural parameters such as crystallite size and dislocation density, with the broadening of XRD peaks induced by stress of the crystalline structure during deformation \cite{Chang,Groma}. Williamson and Hall \cite{wh53} suggested that the two contributions to the broadening can be represented as
\beq
\Delta K =  \frac{0.9}{D} + {\Delta K}^D
\eeq
where $\Delta K= 2 \cos \theta (\Delta \theta ) / \lambda$ is the full width at half maximum (in radians), $\theta$ is the diffraction angle, $\lambda$ is the wavelength, $D$ is the crystallite size and ${\Delta K}^D$ is the strain contribution to line broadening. The Williamson-Hall plot of $\Delta K$ as a function of the scattering vector $K = 2 \sin \theta /\lambda$ was modified by Ung\'ar and Borb\'ely \cite{re9996}, replacing $K$ or $K^2$  by $K\bar{C}^{1/2}$ or$K^2\bar{C}$, respectively. These latter procedures, called `modified' Williamson-Hall method, can be used in the case of systems of isotropic particle sizes for a physically correct determination of the apparent size and mean square strain \cite{cheary,gubicza,scardi}. Thus, if dislocations are the main contributors to the residual strain, the modified Williamson-Hall plot can be expressed as:
\begin{equation}
\frac{\left ( \Delta K \right )^2 - \left ( \frac{0.9}{D}\right )^2 }{K^2} = \left ( \frac{\pi M^2 b^2 \Lambda}{2}\right ) \bar{C}_{h00} (1-qH^2)
\label{EcnDK2a}
\end{equation}
where $M$ is a constant depending on the effective outer cut-off radius of dislocations, and $H^2 = (h^2k^2 + h^2+l^2 + k^2l^2)/(h^2+k^2+l^2)^2$ for a cubic crystal system. $\bar{C}_{h00}$ is the contrast factor for Bragg reflection $(h00)$, which determines the contrast factor for Bragg reflection $(hkl)$ introduced by Ung\'ar and co-workers to take into account the influence of the residual strain over the different Bragg reflections \cite{re13,re9999,re9999b}, through $\bar{C} = \bar{C}_{h00} (1-qH^2)$. From the linear regression of the left-hand side of Eqn. (\ref{EcnDK2}) versus $H^2$,  the parameter $q$ can be determined experimentally. Subsequently, the dislocation density $\Lambda$ can also be determined from XRD results.






\end{document}